# Three-dimensional neural network driving self-interference digital holography enables high-fidelity, non-scanning volumetric fluorescence microscopy

TIANLONG MAN,[1,*] YUWEN ZHANG,[1,†] YUCHEN WU,[2,3,†] ZHIQING ZHANG,[2,3] HONGQIANG ZHOU,[1] LIYUN ZHONG,[4] AND YUHONG WAN[1,*]

[1]*School of Physics and Optoelectronic Engineering, Beijing University of Technology, Beijing, 100124, China*
[2]*Institute of Modern Optics, Nankai University, Tianjin, 300459, China*
[3]*Nankai International Advanced Research Institute (Shenzhen Futian), Shenzhen, 518045, China*
[4]*Key Laboratory of Photonics Technology for Integrated Sensing and Communication of Ministry of Education, Guangdong University of Technology, Guangzhou 510006, China*
[†]*These authors contributed equally.*
*[*t.man@bjut.edu.cn, yhongw@bjut.edu.cn](mailto:t.man@bjut.edu.cn)*

**Abstract:** We present a deep learning driven computational approach to overcome the limitations of self-interference digital holography that imposed by inferior axial imaging performances. We demonstrate a 3D deep neural network model can simultaneously suppresses the defocus noise and improves the spatial resolution and signal-to-noise ratio of conventional numerical back-propagation-obtained holographic reconstruction. Compared with existing 2D deep neural networks used for hologram reconstruction, our 3D model exhibits superior performance in enhancing the resolutions along all the three spatial dimensions. As the result, 3D non-scanning volumetric fluorescence microscopy can be achieved, using 2D self-interference hologram as input, without any mechanical and opto-electronic scanning and complicated system calibration. Our method offers a high spatiotemporal resolution 3D imaging approach which can potentially benefit, for example, the visualization of dynamics of cellular structure and measurement of 3D behavior of high-speed flow field.

## 1. INTRODUCTION

Volumetric fluorescence microscopy has shown its indispensable role in uncovering important structural and functional details in the fields of biology and life science. During the last decades, various efforts have been made in order to push the performances of fluorescence microscopy toward the visualization of dynamic events that happened on the spatial scale of single-molecule-level in cells. While inspiring successes have been achieved in breaking the diffraction barrier that limits the spatial resolution of fluorescence microscopy [1–3], it is yet a challenging task to rapid acquire the 3D fluorescence information. The bottleneck problem is that in majority of the volumetric imaging techniques such as confocal, light-sheet and structural illumination microscopy, axial scanning on the sample and/or the laser beams is necessary to acquire the 3D image which, however, limit the imaging throughput and speed. On the other hand, in non-scanning 3D microscopy methods such as light-field microscopy [4], Fresnel incoherent correlation holography (FINCH) [5,6] and coded aperture correlation holography (COACH) [7,8], the entire 3D fluorescence field can be projected onto a few or even a single far-filed 2D intensity distribution. However, those approaches usually suffer from time-consuming system point spread function (PSF) calibration, relative complicated iterative reconstruction algorithm and customized optical elements that increasing the system complexity. Meanwhile, the background noise that introduced by the twin-images, artifacts and out-of-focus images hinder the non-scanning approaches from providing the 3D reconstructions with comparable quality that of the scanning-based methods. The combination of confocal microscopy and FINCH has significantly improved the axial resolution and suppressed the background noise [9]. The implementation of spinning disk even facilitates faster image acquisition than point-by-point confocal scanning [10]. Unfortunately, these methods are yet limited by system complexity, bulk size, and most important, the necessary scanning of beam.

Recently emerged deep-learning-augmented imaging methods [11–17] have shown the power of such numerical approaches in, for example, enhancing the lateral and axial resolution [18], extending the field-of-depth [19], and retrieving the 3D fluorescence information from a single 2D images that acquired in conventional microscopes [20]. Non-scanning 3D microscopy methods that use deep learning to solve the 3D-3D or 2D-3D inverse problems have presented superior performances [21,22]. Specifically, a deep learning network driven by 3D linear shift-variant imaging model has enabled high-resolution volumetric imaging across a wide field-of-view (FOV) in light-field microscopy [4]. A generative adversarial network (GAN) extends the depth-of-field of FINCH [19]. A pre-calibrated unsupervised deep learning network suppresses the out-of-focus artifacts of COACH [23]. FINCH reconstructions of a 3D sample that composed of two 2D objects exhibit suppressed out-of-focus artifacts when deep neural network was used as reconstruction algorithm [24].

Here we address a fundamental problem in non-scanning 3D microscopy, that how the spatial resolutions along all the three dimensions can be elevated, without any time-consuming calibration on the PSFs and sacrificing of imaging speed. We have shown that instead of the usually adopted 2D model, significant improved spatial resolution, SNR and suppressed out-of-focus artifacts are observed after processing the conventional FINCH reconstructions using trained 3D deep neural networks. To demonstrated these, we developed a high magnification, high numerical aperture fluorescence self-interference digital holographic 3D microscope (SIH3M). We then trained 3D deep neural network model generally called SIH3MNet, with the 3D spatial registered datasets including numerical back-propagation-obtained holographic reconstructions and wide-field (WF) microscopy image stacks. We experimentally validated that with inputs of single complex-valued hologram, SIH3MNet provides high-fidelity 3D reconstructions with spatial resolution that is comparable with WF fluorescence microscope, across a large FOV of ~40 μm × 178 μm × 178 μm ($z, x, y$) which is only limited by the objective lens. We quantified the performances of SIH3MNets to reconstruct fluorescence particles when different structures including 3D Unet, Pseudo 3D Unet and 3D GAN are used.

## 2. METHODS

### A. SIH3M optical system

As shown in Fig. 1, SIH3M is a homebuilt fluorescence self-interference invert-microscope. The excitation light was provided by a 457 nm solid state laser (85-BLT-605, Melles Griot). For the purpose of spatial filtering the laser was coupled into a single-mode fiber. The output light from the fiber was first collimated and then guided into the system using couple of mirrors. The beam diameter of the excitation laser was expanded using a 4$f$ telescope system. To wide-field epi-illuminate the sample, the expanded beam was then reflected by a dichroic mirror (MD498, Thorlabs) and focused toward the back focal plane of the objective lens (60 ×, NA = 0.95, Nikon) by a tube lens (200 mm focal length, AC254-200A, Thorlabs). The sample-emitted fluorescent was collected by the objective lens, separated from the excitation laser by the dichroic mirror, and then filtered by a band pass emission filter (MF525-39, Thorlabs). A spatial light modulator (SLM, 1920 × 1080 pixels, 6.4 μm pixel size, LETO 3.0, Holoeye) was recruited on the conjugated pupil plane. A quadratic phase mask (focal length = 1000 mm) was displayed on the SLM as a beam-splitting element. Together with an achromatic doublet lens (150 mm focal length, AC254-150A, Thorlabs), these elements provide two beams with different wavefront curvatures and orthogonal polarization states before the image detector.

After passing through a linear polarizer (was not shown in Fig. 1), the interference patterns (holograms) that generated by those two beams are captured by the EMCCD camera (13 μm pixel size, iXon Ultra 888, Andor). A single-axis piezo-z-scanning stage (P-736 ZR2S, PI) was used for capturing the 3D wide-field image stacks. SIH3M provides an overall magnification of ~75 × across an area of ~178 μm in lateral ($x$, $y$) direction, with a ~174 nm pixel size. The FOV along axial ($z$) direction depends on the numerical propagation distances that used for 3D reconstruction. The hardware is controlled using μManager.

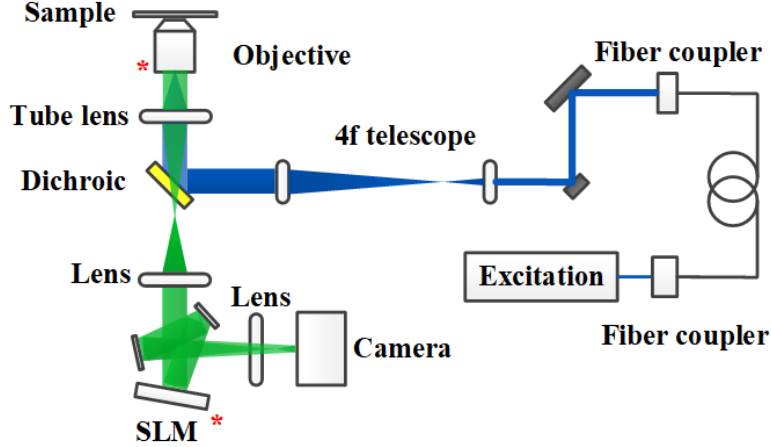

Fig. 1. Schematic of the SIH3M optical system. The asterisks indicate the optical conjugated planes.

## B. SIH3MNet

To enable high quality 3D reconstruction from SIH3M-measured holograms, deep learning models called SIH3MNet is implemented. Each SIH3M hologram includes incoherent composition of Fresnel-zone-plate-like point spread holograms to encode 2D projection information about the 3D object. Since the 2D-3D object reconstruction problem is ill-posed, it is challenging yet to faithfully retrieve the 3D image of object in SIH3M with the presenting of crosstalk artifacts. To address this challenge, SIH3MNet combines two modules to solve the highly ill-posed 3D holographic reconstruction inverse problem, including numerical refocusing and 3D enhancement, as illustrated in Fig. 2.

The first module projects a SIH3M 2D hologram into a 3D volume within object space. This task is performed by propagating the transmitted complex field of the hologram by serial distances along $z$ using the angular spectrum method, where $Q$ is the free space transformation matrix [22]. The intensity of the numerical refocused volume is then fed to the second module "3D enhancement" to remove the artifacts and enhance the spatial resolution and SNR. Here we use the supervised-learning-based SIH3MNet driven by well-registered experimental data pairs because of its inherent advantage in providing high-fidelity outputs, though numerous unsupervised models have been proposed and applied in different optical imaging modalities. Unet-based structure as shown in Fig. 2 is implemented for the 3D image enhancement task.

Four holograms $I_1$, $I_2$, $I_3$, and $I_4$ with different phase shifting values were captured and combined as $H = (I_1-I_3) - j(I_2-I_4)$ to suppress the back-ground noise and twin-images in the 3D reconstructions [6,25]. In the experiments, we measured SIH3M and WF images of a sample with 500 nm diameter fluorescence particles that embedded in agarose gel. 3D WF image stack is captured by mechanical scanning the sample along $z$ using the piezo stage. For the 3D image registration, mismatches in the magnification and voxel size between the SIH3M reconstruction and WF image stacks along the axial direction is first calibrated. This is implemented by using a depth-dependent-nonlinear-spaced reconstruction distances that obtained by manual aligning the intensity of the holographic 3D reconstructed field with the referenced wide-field image stack of the same sample. High accuracy image registration is then implemented using Fijiyama

(Fiji) to align the two datasets along both the lateral and axial directions. Spatial drifts between the phase-shifted holograms and slices from 3D WF image stack are calibrated in Register Virtual Stack (Fiji), if required.

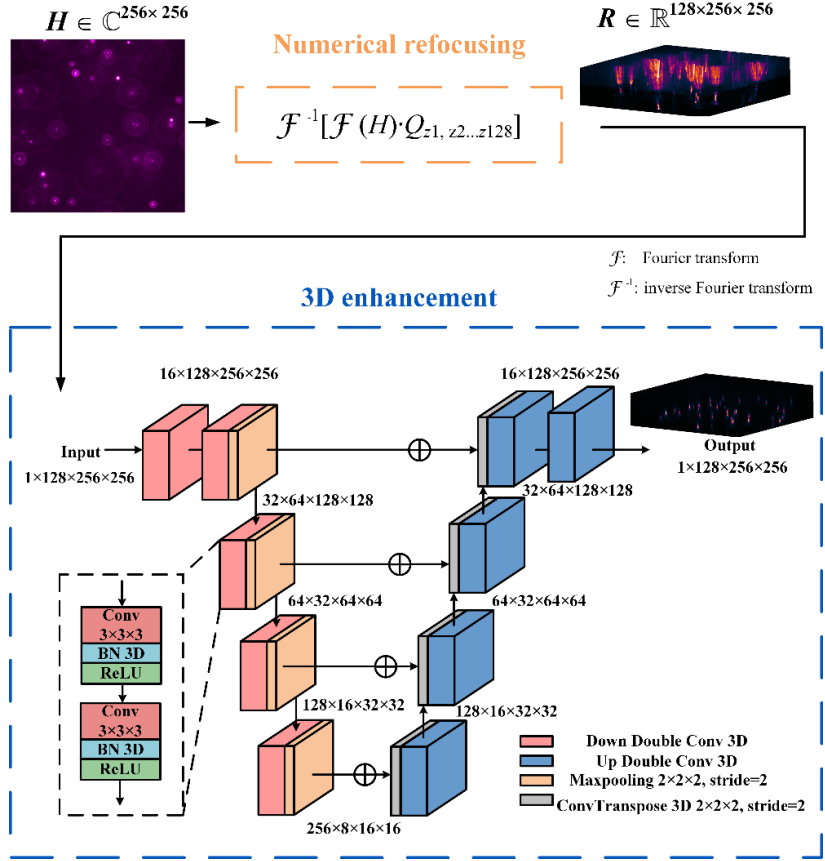

Fig. 2. The model architecture of SIH3MNet.

## 3. RESULTS

Both the aligned SIH3M reconstruction and WF image stack have a data size of 201 × 512 × 512 pixels ($z$, $x$, $y$), with a voxel size (projected into the sample space) of 200 nm × 348 nm × 348 nm. To generate the training data, the image stacks are then resized to 128 × 256 × 256 pixels, corresponding to a voxel size of 314 nm × 696 nm × 696 nm. The network model is optimized by Adam optimizer. The total number of training epochs is set to 200 and we selected the best model with the lowest validation loss. Dataset containing 180 groups of 3D image stacks (with image rotations of 0, 90, 180 or 270 degrees that applied on 45 groups of raw data) are used for network training.

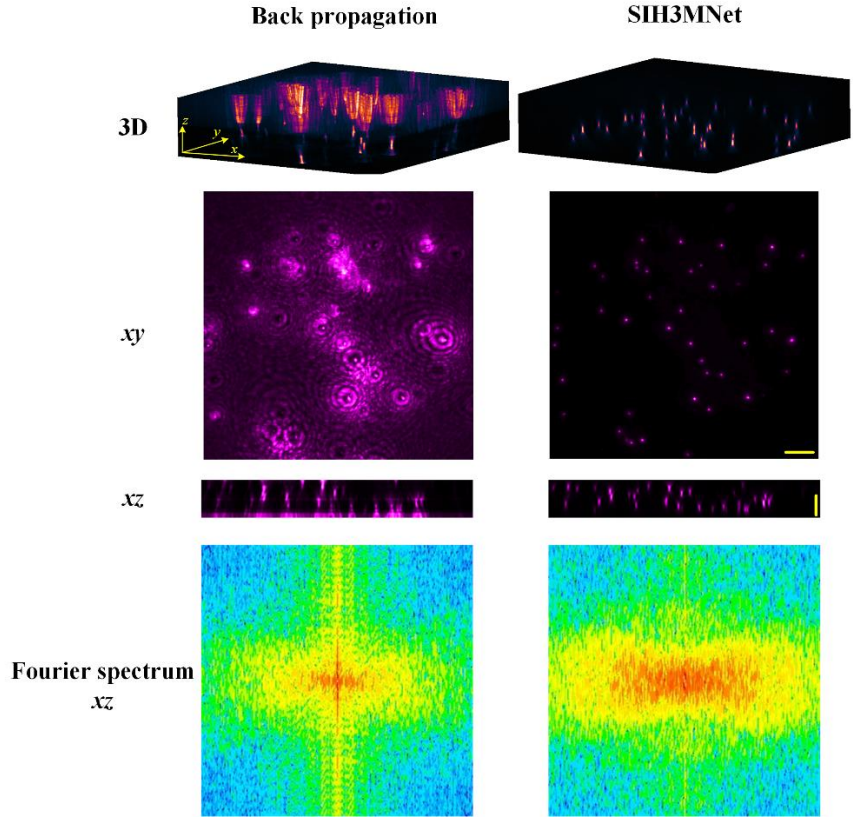

Fig. 3. Holographic nonscanning 3D reconstructions obtained via conventional back propagation method and SIH3MNet. 3D reconstructed images and the corresponding *xy* and *xz* projections (max intensity projections) are presented. Scale bar, 17 μm for *xy* and 20 μm for *z*.

In the testing phase, a batch of test SIH3M reconstructions (128 × 256 × 256 pixels) is fed to the network. The initial learning rate is 1e-4, and the decay strategy of StepLR is adopted to reduce the learning rate by 0.8 times every 10 epochs. We used a mean-squared-error-based loss function. All of the calculations were performed on a desktop with an Intel(R) Core (TM) i5-12400K CPU @ 2.5 GHz, 32 GB of RAM, and an NVIDIA GeForce RTX 4060 Ti graphic card with 16 GB of VARM. The inference time for a full FOV of 40 μm × 178 μm × 178 μm is ~9 s. As shown in Fig. 3, the SIH3M reconstruction obtained by back propagation method is severely degraded by the out-of-focus artifacts. The SIH3MNet output, on the other hand, exhibits significant suppressed artifacts, with an improved resolution and SNR along all the three spatial dimensions. These can be observed clearly when comparing the spatial frequency distribution of the *xz* projections of the 3D reconstructions, since the SIH3MNet output preserves a wider distribution in Fourier domain. The reconstruction performance of SIH3MNet is also quantified in Fig. 4 using the full-width-half-maximum (FWHM).

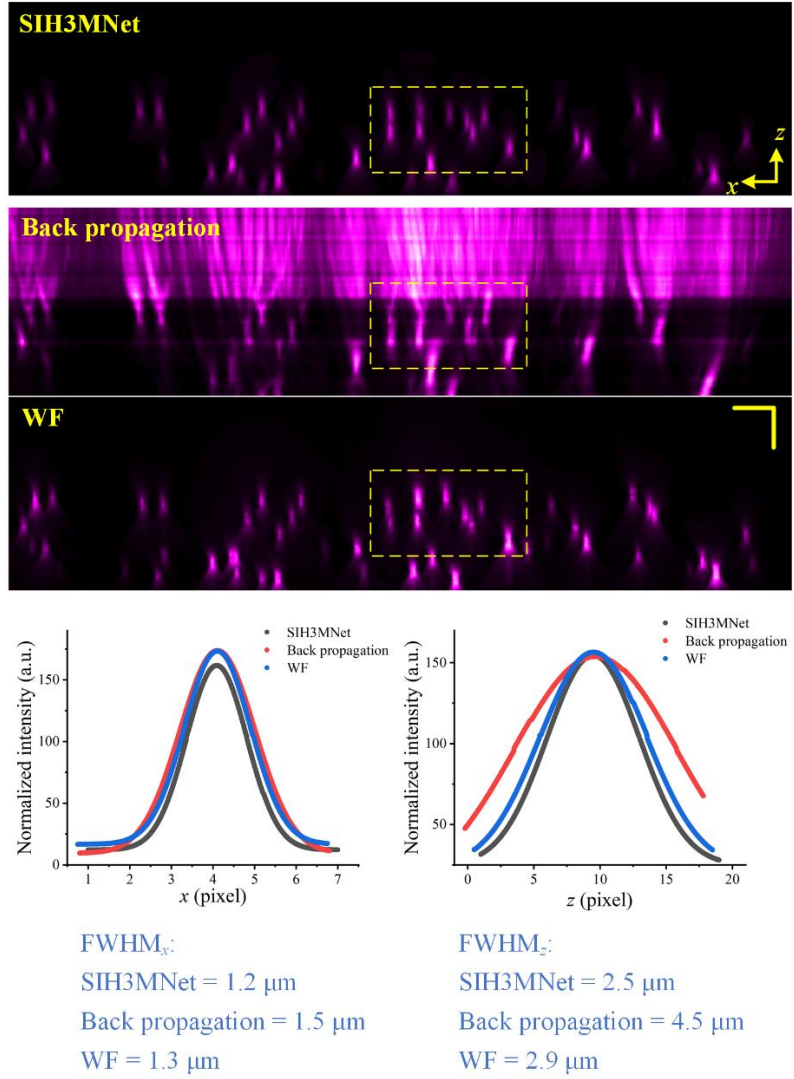

Fig. 4. Quantitative comparison on the spatial resolution of SIH3MNet output, back propagation reconstruction and WF images. Averaged FWHM in the areas that indicated by the yellow boxes are calculated. Scale bar, 8.5 μm for *x* and 8 μm for *z*.

For the experimental results in Fig. 4, we calculated the averaged FWHM of the 3D images of the individual particle within the areas that indicated by the yellow boxes. Simultaneous improvements on both the lateral and axial resolution are obtained in SIH3MNet outputs. The 1.8-fold squeezed $FWHM_z$ has demonstrated the success of our SIH3MNet in elevating one of the major limitations of self-interference holography, i.e., the poor axial imaging performances. Surprisingly, the SIH3MNet even offers a slightly better performances then the ground truth that used in the network training (WF images). These can be the potential results of the partially suppression on the spherical-aberration-introduced artifacts which can be observed when comparing the SIH3MNet output and WF images. All the FWHMs are extended when comparing with their theoretical value (objective lens with NA = 0.95), because of the relatively large voxel size of the network training datasets.

The reconstruction performances of SIH3MNets with different network structure including Pseudo 3D Unet [26] and 3D CycleGAN [27,28] are investigated. As shown in Fig. 5, Pseudo

3D Unet provides slightly extended FWHMs, while the reconstructed images are distorted for the particles that closed to the boundary of the FOV. With strong back ground noise and image distortions, 3D CycleGAN failed in exporting comparable results to that of 3D Unet in our experiments.

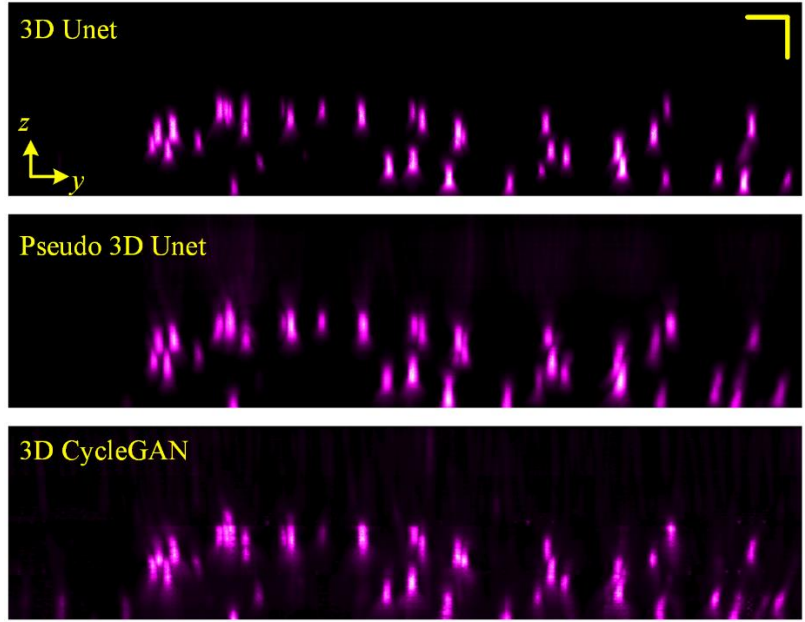

Fig. 5. SIH3MNet outputs when different network structures are implemented. Scale bar, 8.5 μm for *x* and 8 μm for *z*.

## 4. CONCLUSION

In conclusion, we addressed the fundamental 2D-3D holographic reconstruction problem in self-interference digital holography. We have demonstrated that with a single complex valued hologram as input, a trained 3D deep neural networks can evaluate simultaneously the SNR and 3D spatial resolutions of the conventional back-propagation-obtained reconstructions. In the imaging experiments on 3D distributed fluorescent particles, our SIH3MNet achieved a 1.8-fold improved axial resolution, without any sacrificing of imaging speed. SIH3MNets with different network structures then 3D Unet are investigated. However, in our experiments both the Pseudo 3D Unet and 3D CycleGAN fail in providing better reconstruction performance, although in principle the previous one holds faster operation speed and accurate image registrations are not required for the later one. One potential reason is CycleGAN model needs larger dataset to provide better performance. The proposed method offers a superior non-scanning 3D computational imaging technique where only 2D self-interference hologram is required to achieve a high-quality 3D image reconstruction. We will focus our future direction on the validation of SIH3MNet on more complex sample such as fluorescent-labelled cells. Another interesting work toward the self-supervised learning of SIH3MNet, which can be achieved by designing deep neural networks that driven by physical model of the SIH3M imaging system.

**Funding.** National Natural Science Foundation of China (61575009), Natural Science Foundation of Beijing Municipality (4182016), Beijing Municipal Natural Science Foundation (3222001), Applied Basic Research Fund of the School of Physics and Optoelectronic Engineering, Beijing University of Technology (ABRFSPOE05), Young Elite Scientist

Sponsorship Program by BAST (BYESS2023066), Tianjin Key Laboratory of Micro-scale Optical Information Science and Technology.

**Acknowledgment.** We acknowledge the helps of Prof. Y. Qin and Dr. S Wan, Nanyang Normal University, for their help in demonstrating the validity of the method via simulations (results are not shown here).

**Disclosures.** The authors declare no conflicts of interest.

**Data Availability.** The datasets used and/or analyzed during the current study are available from the corresponding author on reasonable request.